\documentstyle[aps,amssymb,graphicx]{revtex}
\twocolumn

\newcommand{\andy}[1]{}

\begin{document}
\title{Finding optimal strategies for minimum-error quantum-state 
discrimination}
\author{M. Je\v{z}ek, J. \v{R}eh\'{a}\v{c}ek\thanks{e-mail:
rehacek@phoenix.inf.upol.cz}, 
and J. Fiur\'{a}\v{s}ek}
\address{Department of Optics, Palacky University, 
17. listopadu 50, 77200 Olomouc,
Czech Republic}
\date{\today}
\maketitle

\begin{abstract}
We propose a numerical algorithm for finding
optimal measurements for quantum-state discrimination.
The theory of the semidefinite programming provides
a simple check of the optimality of the numerically 
obtained results. With the help of our algorithm we
calculate the minimum attainable error rate of 
a device discriminating among three particularly chosen 
non-symmetric qubit states.
\end{abstract}

\vspace{10pt}

Nonorthogonality of quantum states is one of the basic features of
quantum mechanics. Its deep consequences are reflected in all 
quantum protocols. For instance, it is well known that 
perfect decisions between  two nonorthogonal states 
cannot be made. This has important implications for the 
information processing at the microscopic level since 
it sets a limit on the amount of information that can
be encoded into a quantum system. 
Although perfect decisions between nonorthogonal 
quantum states are impossible, it is of importance
to study measurement schemes performing
this task in the optimum, though imperfect, way.

Two conceptually different models of decision tasks have been 
studied. The first one is based on the minimization of the Bayesian 
cost function, which is nothing but a generalized error rate
\cite{helstrom}. In the special case of linearly independent 
pure states, the second model --- unambiguous discrimination
of quantum states --- makes an interesting alternative. 
The latter scheme combines the error-less discrimination 
with a certain fraction of inconclusive results 
\cite{ivanovic,dieks,peres}. 

Ambiguous as well as unambiguous discrimination schemes
have been intensively studied over the past few years. 
In consequence, the optimal measurements distinguishing 
between pair, trine, tetrad states, and linearly independent 
symmetric states are now well understood
\cite{yuen,Chefles_Barnett_1997,Chefles_1998,Chefles_Barnett_1998,%
Phillips_Barnett_Pegg_1998,sasaki,Chefles_2000,walgate,Virmani_et_al_2001,%
Zhang_et_al_2001,Barnett_2001,Chefles_2001,barnett}. 
Many of the theoretically discovered optimal devices have 
already been realized experimentally, mainly with polarized 
light \cite{Huttner_et_al_1996,Barnett_Riis_1997,%
Clarke_Chefles_Barnett_Riis_2001,Clarke_et_al_2001}.
As an example of the practical importance of the optimal decision
schemes let us mention their use for the eavesdropping on quantum   
cryptosystems \cite{Ekert_et_al_1994,Dusek_et_al_2000}.

The purpose of this paper is to develop universal
method for optimizing ambiguous discrimination between 
generic quantum states.

Assume that Alice sets up $M$ different sources of 
quantum systems living in $p$-dimensional Hilbert space.
The complete quantum-mechanical description of each source is
provided by its density matrix. Alice chooses 
one of the sources at random using a chance device
and sends the generated quantum system to Bob.
Bob is also given $M$ numbers $\{\xi_i\}$ specifying 
probabilities that $i$-th source is selected
by the chance device. Bob is then required to 
tell which of the $M$ sources $\{\rho_i\}$ 
generated the quantum system he had obtained from
Alice. In doing this he should make as few mistakes
as possible.

It is well known \cite{helstrom} that
each Bob's strategy can be described in terms of an 
$M$-component probability operator measure (POM) $\{\Pi_j\}$,
$0<\Pi_i<1$, $\sum_j \Pi_j=1$. Each POM element 
corresponds to one output channel of Bob's discriminating 
apparatus. The probability that Bob points his finger at the
$k$-th source while the true source is $j$ is given
by the trace rule: $P(k|j)={\rm Tr}\rho_j\Pi_k$.
Taking the prior information into account, the average 
probability of Bob's success in repeated experiments is
\andy{probab}
\begin{equation}\label{probab}
P_s=\sum_{j=1}^M\xi_j{\rm Tr}\rho_j\Pi_j.
\end{equation}
Since the objective is to keep Bob's error rate as low as
possible we should maximize this number over the set of all
$M$-component POMs. In compact form the problem reads:
\andy{problem}
\begin{equation}\label{problem}
\begin{array}{l}
\rm{maximize}\; P_s\;\mbox{subject to constraints}\\[2pt]
\Pi_j\ge 0, \quad j=1\ldots M,\\[3pt]
\sum_j\Pi_j=1.
\end{array}
\end{equation}
Unfortunately, attacking this problem by analytical means 
has chance to succeed only in the simplest cases ($M=2$)
\cite{helstrom}, or cases with symmetric or 
linearly independent states 
\cite{yuen,sasaki,Barnett_2001,barnett,ban}.
In most situations one must resort to numerical methods.
In the following we will use the calculus of variations
to derive a simple iterative algorithm that provides a 
convenient way of dealing with the problem (\ref{problem}).
This approach has already found its use in the optimization 
of teleportation protocols \cite{teleportation}
and maximum-likelihood estimation of quantum measurements
\cite{ml-measurements}.
We are going to seek the global maximum of the success
functional $P_s$ subject to the constraints given in 
Eq.~(\ref{problem}).
To take care of the first constraint we will decompose
the POM elements as follows
$\Pi_j=A_j^{\dag}A_j,\; j=1\ldots M$.
The other constraint (completeness) can be  
incorporated into our model using the method of uncertain 
Lagrange  multipliers. Putting all things together, 
the functional to be  maximized becomes 
\andy{funct}
\begin{equation}\label{funct}
{\cal L}=\sum_j \xi_j\rm{Tr}\{\rho_j A_j^{\dag}A_j\}-
\rm{Tr}\{\lambda\sum_j A_j^{\dag}A_j\},
\end{equation}
where $\lambda$ is a Hermitian Lagrange operator. This expression
is now to be varied with respect to $M$ independent variables
$A_j$ to yield a necessary condition for the extremal 
point in the form of a set of $M$ extremal 
equations for the unknown POM elements:
$\xi_j\rho_j\Pi_j=\lambda\Pi_j,\; j=1\ldots M$. For our 
purposes it is advantageous to bring these equations 
to an explicitly positive semidefinite form,
\andy{extremal}
\begin{equation}\label{extremal}
\Pi_j=\xi_j^2\lambda^{-1}\rho_j\Pi_j\rho_j\lambda^{-1}, \quad j=1\ldots M.
\end{equation}
Lagrange operator $\lambda$ is obtained by summing 
Eq.~(\ref{extremal}) over $j$,
\andy{lagrange}
\begin{equation}\label{lagrange}
\lambda=\left(\sum_j\xi_j\rho_j\Pi_j\rho_j\right)^{1/2}.
\end{equation}
The iterative algorithm comprised of the $M+1$ equations 
(\ref{extremal}) and (\ref{lagrange}) is the main formal 
result  of this paper.
One usually starts from some ``unbiased'' trial
POM $\{\Pi^0_j\}$. After plugging it in Eq.~(\ref{lagrange})
the first guess of the Lagrange operator $\lambda$ is obtained.
This operator is, in turn, used in Eq.~(\ref{extremal}) to get
the first correction to the initial-guess strategy
$\{\Pi^0_j\}$ \cite{pretty-good}.
The procedure gets repeated, until, eventually, a stationary
point is attained. Notice that both the positivity and completeness
of the initial POM are preserved in the course of iterating.

Since equations (\ref{extremal}) and  (\ref{lagrange})
represent only a necessary condition for the extreme,
one should always check the optimality of the stationary
point.  In the following we
will make use of the theory of the semidefinite
programming (SDP) \cite{semidef} to derive a
criterion of the optimality of the iteratively obtained POM,
which turns out to be the well-known Helstrom
condition \cite{helstrom}.
SDP theory also provides alternative means of solving
the problem (\ref{problem}) numerically.

Recently, it has been pointed out \cite{semidef-appl}
that many problems of the quantum-information processing
can be formulated as SDP problems. For instance,
let us compare our original problem (\ref{problem})
to the SDP dual problem that is defined as follows:
\andy{semi-dual}
\begin{equation}\label{semi-dual}
\begin{array}{l}
{\rm maximize}\; -{\rm Tr} F_0 Z,\\[2pt]
Z\ge 0,\\[3pt]
{\rm Tr} F_i Z=c_i,\quad i=1\ldots m,
\end{array}
\end{equation}
where data are $m+1$ Hermitian matrices $F_i$ and a complex vector 
$c\in {\mathbb C}^m$, and $Z$ is a Hermitian variable.
As can be easily checked, our problem (\ref{problem})
reduces to a dual SDP problem upon the following 
substitutions: 
\andy{correspond}
\begin{eqnarray}\label{correspond}
F_0&=&-\bigoplus_{j=1}^m \xi_j\rho_j,\quad 
Z=\bigoplus_{j=1}^m \Pi_j,\nonumber\\
F_i&=&\bigoplus_{j=1}^m\Gamma_i,\quad
c_i={\rm Tr}\Gamma_i, \quad i=1\ldots p^2.
\end{eqnarray}
Here operators $\{\Gamma_i, i=1\ldots p^2\}$ comprise
an orthonormal operator basis in the $p^2$-dimensional space of 
Hermitian operators acting in the Hilbert space of our problem: 
${\rm Tr}\Gamma_j\Gamma_k=\delta_{jk},\; j,k=1\ldots p^2$.
For simplicity, let us take $\Gamma_1$ proportional to the
unity operator, then all $c_i$ apart from $c_1$ vanish.

An important point is that there exists a primal problem
associated with the dual one,
\andy{prim-feasib}
\begin{equation}\label{semi-primal}
\begin{array}{l}
{\rm minimize} \; c^T x, \\
F(x)=F_0+\sum_i x_iF_i\ge 0.
\end{array}
\end{equation}
Here data $F_i$ and $c_i$ are the same as in
Eq.~(\ref{semi-dual}), and vector $x$ is now the variable.

The advantage of the SDP formulation of the quantum-state
discrimination problem is that there are strong
numerical tools designed for solving SDP problems.
In particular, these methods are guaranteed to
converge to the real solution. This might become important
when the iterative algorithm derived above 
encounters convergence problems.    

A primal (dual) SDP problem is called ``strictly feasible''
if there exists $x$ ($Z$) satisfying the constraints
in Eq.~(\ref{semi-primal}) [Eq.~(\ref{semi-dual})] with sharp
inequalities. One can easily check that both the primal and
dual problems associated with the quantum-state discrimination
problem are strictly feasible. Hence we can use a powerful result
of SDP theory saying that in this case, $x$ is optimal
if and only if $x$ is primal feasible and there is a dual feasible
$Z$ such that
\andy{dual-slackness}
\begin{equation}\label{compl-slackness}
Z F(x)=0.
\end{equation}
This condition is called the complementary slackness condition. 
Now, taking $x_i$ to be the coordinates of 
the Lagrange operator $\lambda$ in $\Gamma_i$ basis, 
$x_i={\rm Tr}\lambda \Gamma_i$, $i=1\ldots p^2$,
the complementary slackness condition is seen to be 
equivalent to the extremal equation.
Since the dual feasibility of the iteratively obtained POM
elements is guaranteed by construction, our extremal
equation becomes a necessary and sufficient condition
on the maximum of $P_s$ once the positive semidefiniteness
of $F(x)$ is verified. In terms of states and POM elements
this latter condition reads:
\begin{equation}\label{criterion}
\lambda-\xi_j\rho_j\ge 0,\quad j=1\ldots M
\end{equation}
One perceives that complementary slackness condition
(\ref{compl-slackness}) together with the criterion of 
optimality (\ref{criterion}) are nothing else than 
the well-known Helstrom equations \cite{helstrom}
for POM maximizing the success probability (\ref{probab}).

\begin{figure}
\centerline{
\includegraphics[width=0.5\columnwidth]{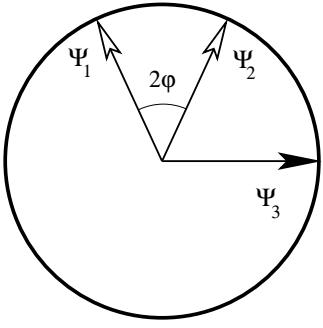}
}
\vspace*{5pt}
\caption{A cut through the Bloch sphere showing the states to be 
discriminated.}
\label{fig-vectors}
\end{figure}

Let us illustrate the utility of our algorithm on a 
simple, albeit nontrivial example of discriminating between  
three non-symmetric coplanar qubit states.
The geometry of this problem is shown in Fig.~\ref{fig-vectors}.
$\Psi_1$ and $\Psi_2$ taken to be equal-prior states, $\xi_1=\xi_2=\xi/2$,
symmetrically placed around the $z$ axis; 
the third state lies in the direction of $x$ or $y$. 
A similar configuration (with $\Psi_3$ lying along $z$) has recently been 
investigated by Andersson {\em et al.} \cite{barnett}. 
Exploiting the mirror symmetry of their problem the authors
derived analytic expressions for POMs minimizing the average error 
rate. For a given angle $\varphi$ the optimum POM turned out
to have two or three nonzero elements depending on 
the amount of the prior information $\xi$. 

\begin{figure}
\centerline{
\includegraphics[width=0.9\columnwidth]{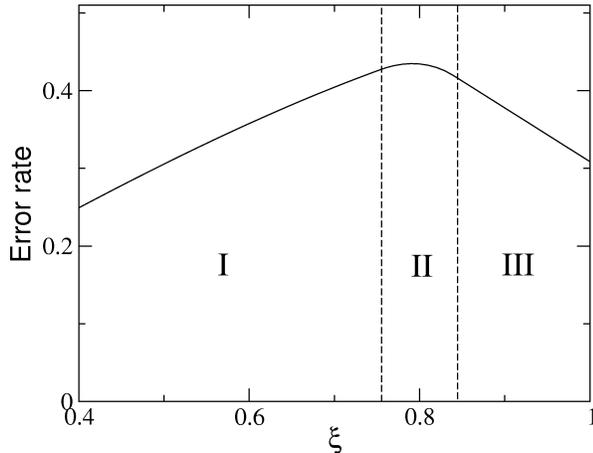}
}
\caption{Average error rate ($1-P_s$) in dependence on Bob's prior 
information $\xi$; $\varphi=\pi/16$. Regions I, II, and III are regions 
where the optimum discriminating device has two, three, and two output 
channels, respectively.}
\label{fig-error}
\end{figure}

Our problem is a bit more complicated one due to the lack 
of the mirror symmetry. Let us see whether the transition from
the mirror-symmetric configuration to a non-symmetric one
has some influence on the qualitative behavior of the optimal POMs.
Minimal error rates calculated using the proposed iterative procedure
[Eqs.~(\ref{extremal}) and (\ref{lagrange})] for the fixed angle of
$\varphi=\pi/16$ are summarized in Fig.~\ref{fig-error}.
The conclusions that can be drawn from the numerical results
partly coincide with that of Ref.~\cite{barnett}. For large $\xi$
(region III) the optimum strategy consist in the optimal discrimination 
between states $\Psi_1$ and $\Psi_2$.
When $\xi$ becomes smaller than a certain $\varphi$-dependent 
threshold (region II), state $\Psi_3$ can no longer be ignored and 
the optimum POM has three nonzero elements.
Simple calculation yields 
\andy{boundary}
\begin{equation}\label{boundary}
\xi_{\rm II,III}=\frac{1}{1+\sin\varphi\cos\varphi}
\end{equation}
for the threshold value of the prior.
However, when $\xi$ becomes still smaller (region I), the optimum
POM will eventually become a two-element POM again 
-- the optimal strategy now being the optimal 
discrimination between states $\Psi_1$ and $\Psi_3$.
This last regime is absent in the mirror-symmetric 
case. The transition between regions I and II is governed 
by a much more complicated expression than Eq.~(\ref{boundary}),
and will not be given here. 
We will close the example noting that 
already a few iterations are enough to determine the optimum
discriminating device to the precision the elements of the 
realistic experimental setup can be controlled with in the 
laboratory.

In this paper we derived a simple iterative algorithm
for finding optimal devices for quantum-state discrimination.
Utility of our procedure was illustrated on a non-trivial example 
of discriminating between three non-symmetric states.
From the mathematical point of view, the problem of quantum-state 
discrimination is a problem of the semidefinite programming.
Such correspondence is a good news since there exist robust 
numerical tools designed to deal with SDP problems.
These can substitute our iterative algorithm in the very few
exceptional cases (if there are any) where our procedure might 
suffer from the convergence problems.

This work was supported by grant No. LN00A015 and
project CEZ:J14/98 ``Wave and particle optics''   
of the Czech Ministry of Education.


\begin{thebibliography}{9}
\bibitem{pretty-good}
It is interesting to note that when the initial POM is chosen
to be the maximally ignorant one, $\Pi^0_j=1/M,\;j=1\ldots M$,
the first correction is, in fact, the ``pretty good'' measurement 
introduced by P. Hausladen and W.K. Wootters, J. Mod. Opt. {\bf 41},
2385 (1994). Interestingly enough, this measurement is known
to be optimal in certain cases \cite{ban}.
\end{thebibliography}
\end{document}